# Evidence for hidden fermion that triggers high-temperature superconductivity in cuprates

Shiro Sakai[1,2], Marcello Civelli[3], and Masatoshi Imada[1]

[1]Department of Applied Physics, University of Tokyo, Bunkyo-ku, Tokyo 113-8656, Japan.
[2]Center for Emergent Matter Science, RIKEN, Wako, Saitama 351-0198, Japan.
[3]Laboratoire de Physique des Solides, Université Paris-Sud, CNRS, UMR 8502, F-91405 Orsay Cedex, France.

In superconductors, electrons bound into Cooper pairs conduct a dissipationless current. The strength of the Cooper pairs scales with the value of the critical transition temperature ($T_c$). In cuprate high-$T_c$ superconductors, however, the pairing mechanism is still unexplained. Here we unveil why in the cuprates the Cooper pairs are so strongly bound to work out the extraordinary high $T_c$. From one-to-one correspondence between numerical simulation on a microscopic cuprate model and a simple two-component fermion model, we show that hidden fermions emerge from the strong electron correlation and give birth to the strongly bound Cooper pairs. This mechanism is distinct from a conventional pairing mediated by some bosonic glue, such as phonons in conventional superconductors. The hidden fermions survive even above $T_c$ and generate the strange-metal pseudogap phase. This reveals an unprecedented direct relationship between the pseudogap phase and superconductivity in the cuprates.

In conventional superconductors, phonons were proven to be the glue of the Cooper pairing: As Migdal-Eliashberg (ME) theory[1] had predicted, the frequency($\omega$)-dependent superconducting gap function $\Delta(\omega)$, measured by electronic tunneling or optical reflectivity, conformed to the phonon-frequency distribution[2-5]. Further evidence of the phonon glue was that in strong-coupling superconductors, e.g., Pb, prominent peaks in Im$\Delta$ (Fig.1a) associated with specific phonons exist, which enhance the pairing gap Re$\Delta(\omega=0)$ through the Kramers-Kroenig relation. Here, Re$\Delta(\omega=0)$ is the measure of the Cooper pair strength.

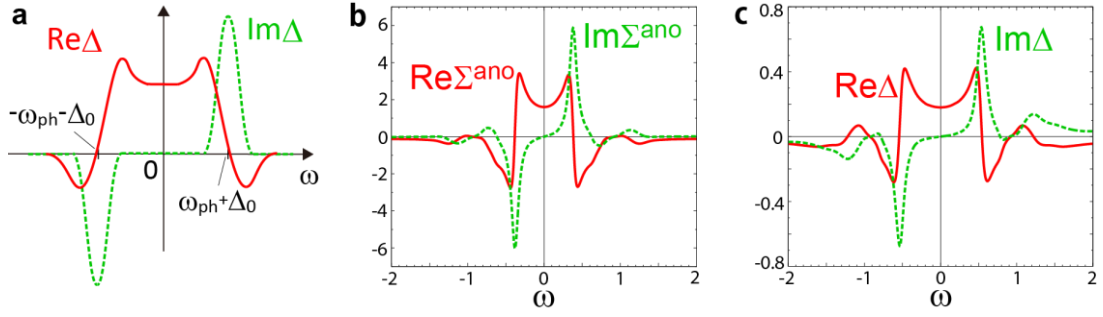

**Figure 1 | ω dependence of the gap function. a,** A scheme for conventional superconductors. $\omega_{ph}$ is the phonon frequency and $\Delta_0$ is the gap-edge energy[2]. **b,** Anomalous self-energy at $\mathbf{k}=\mathbf{k}_{AN}=(\pi,0)$ calculated with the CDMFT for the 2D Hubbard model[10-13]. The hole doping $x$=0.05 and temperature $T$=0.01. **c,** Gap function for the same parameters.

In high-$T_c$ cuprates, on the other hand, the glue has not been identified yet. It was proposed that a glue may even be unnecessary[6] in the strongly-correlated state close to the Mott insulator. In this latter case, the ME theory would not be applicable. Nevertheless the ω and momentum (**k**) dependences of the gap function $\Delta(\mathbf{k},\omega)$ still provide important information about the superconducting mechanism.

It has been well established that $\Delta(\mathbf{k},\omega)$ of the cuprates shows a *d*-wave-like **k** dependence, differently from conventional superconductors. Much less attention has, however, been paid to its ω dependence.

The development of cluster dynamical mean-field theory (CDMFT)[7,8] has enabled to microscopically calculate $\Delta(\mathbf{k},\omega)$. On top of the *d*-wave **k** dependence[7,9], CDMFT has revealed that a quantity called the anomalous self-energy $\Sigma^{ano}(\mathbf{k},\omega)$, closely related to $\Delta(\mathbf{k},\omega)$ (as detailed in Eq.3 below), has prominent peaks in ω dependence (Fig.1b), analogous to $\Delta$ of the conventional strong-coupling superconductors[10-13]. The origin of the peaks in Im$\Sigma^{ano}$ was attributed to some bosonic glue, including spin fluctuations which would play the role of phonons in the conventional superconductors[10,11].

CDMFT has also reproduced anomalous metallic behaviors experimentally observed in the pseudogap phase above $T_c$.[7,14] Its relationship with the superconductivity is a central open issue[13,15-18].

In this report, we reveal that the prominent peak in Im$\Sigma^{ano}$ (or Im$\Delta$), which is at the origin of the strong Cooper pairing, is generated by hidden fermions, and hereby establish a fermionic strong-pairing mechanism distinct from the conventional bosonic glue. We also demonstrate that the pseudogap arises from the same hidden fermions.

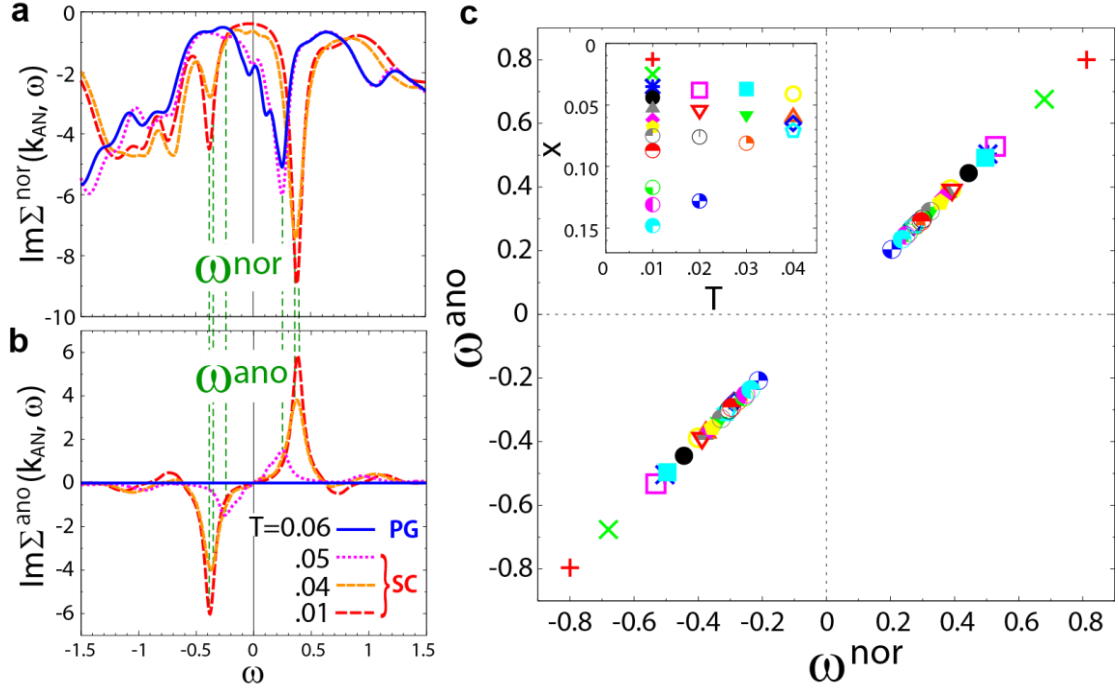

**Figure 2 | The peak-to-peak correspondence between Im$\Sigma^{nor}$ and Im$\Sigma^{ano}$. a,** Im$\Sigma^{nor}$ calculated with the CDMFT at **k**=**k**$_{AN}$=$(\pi,0)$ for $t'$=-0.2, $U$=8 and $x$~0.05. **b,** Im$\Sigma^{ano}$ for the same parameters. $T_c$ is found to be in between $T$=0.05 and 0.06. Green dashed lines point to the low-energy peak positions, $\omega^{nor}$ and $\omega^{ano}$. **c,** $\omega^{nor}$ and $\omega^{ano}$ plotted for various $n$ and $T$ (shown in inset) in the superconducting state.

We first show the numerical results obtained with CDMFT for the superconducting phase. Following the previous studies[7-18], we employ the two-dimensional (2D) Hubbard model, a standard model for the high-$T_c$ cuprates consisting of the electronic transfer $t$ (whose Fourier transform to momentum space is the bare dispersion $\varepsilon(\mathbf{k})$) and the Coulomb interaction $U$; see Methods for details. We set the energy unit $t$=1 throughout the paper. The typical CDMFT output is Green's function

$$G(\mathbf{k},\omega)=[\omega-\varepsilon(\mathbf{k})-\Sigma^{nor}(\mathbf{k},\omega)-W(\mathbf{k},\omega)]^{-1} \qquad (1)$$

with

$$W(\mathbf{k},\omega)=\Sigma^{ano}(\mathbf{k},\omega)^2/[\omega+\varepsilon(\mathbf{k})+\Sigma^{nor}(\mathbf{k},-\omega)^*], \qquad (2)$$

as formulated in Methods. The normal self-energy $\Sigma^{nor}$ and the anomalous self-energy $\Sigma^{ano}$ give $\Delta$ through[4]

$$\Delta(\mathbf{k},\omega)=\Sigma^{ano}(\mathbf{k},\omega) / [1-(\Sigma^{nor}(\mathbf{k},\omega)-\Sigma^{nor}(\mathbf{k},-\omega)^*)/2\omega]. \qquad (3)$$

Im$\Delta$ calculated through Eq.3 shows a peak structure (Fig.1c) similar to that of Im$\Sigma^{ano}$.[19]

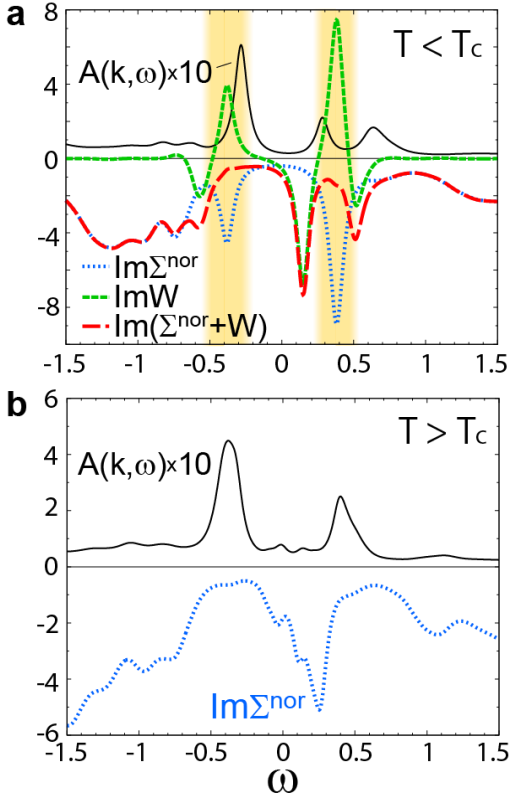

**Figure 3 | Cancellation of the poles. a,** Im$\Sigma^{nor}$ (blue dotted curve), Im$W$ (green dashed curve), their sum (red long-dashed curve), and $A(\mathbf{k},\omega)$ (black solid curve) calculated with the CDMFT at $\mathbf{k}=(\pi,0)$ for $T=0.01$ and x=0.05. **b,** Results at $T=0.06$.

We confirm that the prominent peak in Im$\Delta$ at $\omega \sim 0.5$ contributes more than 80% to the gap

$$\mathrm{Re}\Delta(\mathbf{k}, \omega = 0) = \frac{2}{\pi}\int_0^\infty \frac{\mathrm{Im}\Delta(\mathbf{k},\omega')}{\omega'} d\omega'$$ (see

Supplementary Fig.S1), ensuring that this peak indeed sends $T_c$ soaring. Therefore, the origins of the peaks in Im$\Delta$ (Fig.1c) and associated Im$\Sigma^{ano}$ (Fig.1b) are our primary interest.

Figures 2a and b show Im$\Sigma^{ano}$ and Im$\Sigma^{nor}$ below and above $T_c \sim 0.05$, in the underdoped region (hole doping $x\sim 0.05$). We show data at the antinodal point $\mathbf{k}=\mathbf{k}_{AN}\equiv(\pi,0)$ because of its major contribution to the $d$-wave superconductivity. Remarkably, all the peak energies ($\omega^{ano}$; equivalent to the pole positions at $T=0$) of Im$\Sigma^{ano}$ coincide with those ($\omega^{nor}$) of Im$\Sigma^{nor}$ for $T<T_c$, as plotted in Fig.2c, suggesting the two peaks have the same origin.

In Fig.3a we plot Im$\Sigma^{nor}(\mathbf{k}_{AN},\omega)$, Im$W(\mathbf{k}_{AN},\omega)$ and their sum, which appears in the denominator of $G$ in Eq.1. It reveals that Im$\Sigma^{nor}$ and Im$W$ in the shaded regions not only have identical peak positions but surprisingly cancel out in the sum, making the peaks invisible in the spectral function $A(\mathbf{k},\omega)= -\mathrm{Im}G(\mathbf{k},\omega)/\pi$. This cancellation in $G$ cannot be explained by the bosonic-glue mechanism of the ME theory, that motivates us to introduce a hidden fermion below.

When we enter the normal state above $T_c \sim 0.05$, $\Sigma^{ano}$ vanishes while the peak of Im$\Sigma^{nor}$ continuously evolves into the peak at $\omega \sim 0.25$ for $T=0.06$ (Figs.2a and 3b). This peak of Im$\Sigma^{nor}$ generates the pseudogap[20-24], by suppressing $A(\mathbf{k},\omega)$ for $-0.3<\omega<0.3$ (Fig.3b) because the peak cancellation with $W$ does not occur anymore. This evidences that the pseudogap at $T>T_c$ shares the same origin with the peak of Im$\Sigma^{ano}$ that generates the high-$T_c$ superconductivity below $T_c$.

A closer inspection shows that $\omega^{nor}$ continuously increases as $T$ decreases (Fig.2a) (corresponding to the growing pseudogap) and eventually, at $T=T_c$, it crosses the quasiparticle energy $\omega^{qp}$, which is identified with the peak in $A(\mathbf{k},\omega)$ ($\omega^{qp} \sim 0.3$ in Fig.3a).

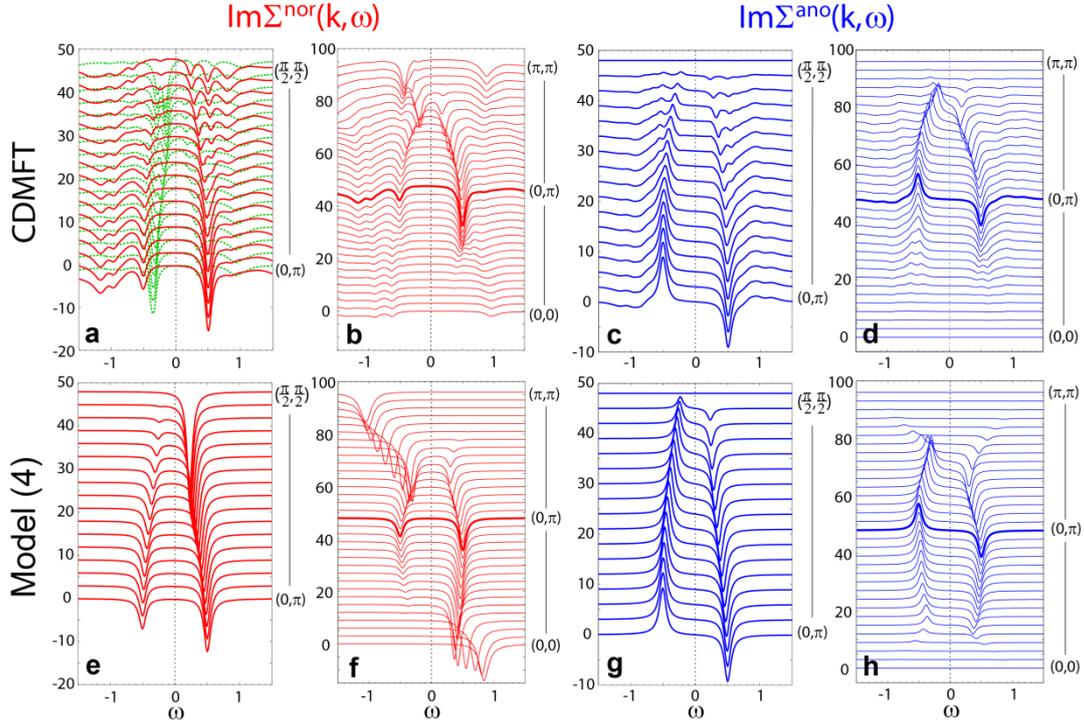

**Figure 4 | Fig.4. k dependence of the normal and anomalous self-energies along $(0,\pi)$-$(\pi/2,\pi/2)$ and $(0,0)$-$(0,\pi)$-$(\pi,\pi)$ lines. a-d,** The CDMFT results for $T=0.01$ and $x=0.04$. Green dashed curves in panel A show $-20 \times A(\mathbf{k},\omega)$. **e-h,** The results for the model (Eq.4), calculated according to Eq.5, at $\mu_f=-0.23$, $D_f^0=0.49$, and $V=1.1$. For clarity, each curve is shifted by 3 along the vertical axis.

Below $T_c$, the relation $\omega^{ano} = \omega^{nor} > \omega^{qp}$ (Fig.3a) enables (i) the quasiparticles (eventually bound into Cooper pairs) to exist around the Fermi level, and (ii) the peak in $\mathrm{Im}\Sigma^{ano}$ to enhance the superconductivity. The crossing of $\omega^{nor}$ and $\omega^{qp}$ appears to determine $T_c$.

To understand physical origin of the mathematically remarkable structure in $\Sigma^{ano}$, $\Sigma^{nor}$ and $\Delta$, we consider a two-component fermion model,

$$H_{\mathrm{TCF}} = \sum_{\mathbf{k}\sigma}\left[\varepsilon_c(\mathbf{k})c^\dagger_{\mathbf{k}\sigma}c_{\mathbf{k}\sigma} + \varepsilon_f(\mathbf{k})f^\dagger_{\mathbf{k}\sigma}f_{\mathbf{k}\sigma} + V(c^\dagger_{\mathbf{k}\sigma}f_{\mathbf{k}\sigma} + f^\dagger_{\mathbf{k}\sigma}c_{\mathbf{k}\sigma})\right] \\ - \sum_{\mathbf{k}}\left[D_c(\mathbf{k})c_{\mathbf{k}\uparrow}c_{-\mathbf{k}\downarrow} + D_f(\mathbf{k})f_{\mathbf{k}\uparrow}f_{-\mathbf{k}\downarrow} + \mathrm{h.c.}\right] \quad (4)$$

and show that the model strikingly reproduces the whole CDMFT results at low energy. Here $c$ represents the quasiparticles and $f$ a hidden fermion-type excitation which hybridizes to $c$ via the coupling $V$. The $V$ term derives from the Hubbard $U$ (see Supplementary Information for details). The $c$ ($f$) bare dispersions and superconducting order parameters are denoted by $\varepsilon_{c(f)}$ and $D_{c(f)}$, respectively. Here, some initial trigger of pairing is assumed by $D_{c(f)}$ which originates from, for instance, some bosonic glue or other excitations. As we will show below, this initial trigger is not the primary origin of the boosting up of $T_c$. By analytically solving the Hamiltonian (Eq.4), the self-energies for $c$ result in

$\Sigma_c^{nor}(\mathbf{k},\omega)=V^2 (\omega+\varepsilon_f(\mathbf{k}))/(\omega^2-\varepsilon_f(\mathbf{k})^2- D_f(\mathbf{k})^2)$,

$\Sigma_c^{ano}(\mathbf{k},\omega)=D_c(\mathbf{k})-V^2 D_f(\mathbf{k})/(\omega^2-\varepsilon_f(\mathbf{k})^2-D_f(\mathbf{k})^2)$ (5)

(see Methods) Notice that poles of $\Sigma_c^{nor}$ and $\Sigma_c^{ano}$ appear at the same energies $\omega^{nor}=\omega^{ano}=\pm\sqrt{\varepsilon_f^2+D_f^2}$, in agreement with the CDMFT result (Fig.2). Moreover, the residues of $W$ and $\Sigma_c^{nor}$ at $\omega=\pm\sqrt{\varepsilon_f^2+D_f^2}$ surprisingly cancel with each other:

$$R_W(\pm\sqrt{\varepsilon_f^2+D_f^2}) \equiv \lim_{\omega\to\pm\sqrt{\varepsilon_f^2+D_f^2}}\left(\omega\mp\sqrt{\varepsilon_f^2+D_f^2}\right)W(\mathbf{k},\omega)=-\frac{V^2}{2}\left(1\pm\frac{\varepsilon_f}{\sqrt{\varepsilon_f^2+D_f^2}}\right)$$
$$=-R_{\Sigma_c^{nor}}(\pm\sqrt{\varepsilon_f^2+D_f^2})$$
(6)

This perfectly explains the cancellation of peaks in Fig.3a.

Furthermore, Eq.5 can be fitted to the CDMFT results (Figs.4a-d). In Figs.4e-h we assumed $\varepsilon_f(\mathbf{k}) = \varepsilon_c(\mathbf{k}+\mathbf{Q}) - \mu_f$ with $\varepsilon_c$ the renormalized dispersion of quasiparticle $c$, $\mu_f$ the chemical potential for the $f$ fermion and $\mathbf{Q}=(\pi,\pi)$ (see Supplementary Information for details). Then the model (Eq.4) reproduces remarkably well the main features (including the dispersion, relative amplitudes between the $\omega>0$ and $\omega<0$ sides, as well as between the normal and anomalous parts) of the CDMFT results. This means that the Hubbard $U$ interaction is well captured by Eq.4 in the low-energy range. Note that in Fig.4a $|\omega^{ano}|$ is always larger than $|\omega^{qp}|$, implying the enhancement of the superconducting gap works at all momenta.

We show in Supplementary Information further physical insights gained by the comparison between the CDMFT results and the solution of the model Eq.4. They include the pole-energy shift between $\Delta$ and $\Sigma^{ano}$, and the distinction between the superconducting gap and pseudogap in terms of different relevant singularities.

All the above identical mathematical structures provide firm evidences for the existence of the hidden fermion that boosts up $T_c$ and generates the pseudogap, in distinction from a solely bosonic-glue scenario.

A candidate for the *f* fermion is the antiferromagnetic counterpart to *c*, as in the mean-field theory of the antiferromagnetic order with the Bragg peak **Q**, where $c_\mathbf{k}$ hybridizes with $c_{\mathbf{k+Q}}$. Another candidate is $\tilde{f}_{i\sigma} \equiv \sqrt{\frac{m}{1-m}} c_{i\sigma} - \sqrt{\frac{1}{m(1-m)}} c_{i\sigma} n_{i\bar{\sigma}}$ with *m* the electron filling per spin. This $\tilde{f}$ is interpreted as the composite fermion proposed in Ref.25, which describes an incoherent part of electrons. $\tilde{f}$ approximately satisfies the anticommutation relations with *c*, and recasts the Hubbard *U* term into a hybridization between *c* and $\tilde{f}$ (see Supplementary Information). The Hubbard model is then mapped to the form of Eq.4. Under the strong antiferromagnetic correlations, $\tilde{f}$ becomes $(-1)^i c_{i\sigma}$ at half filling (*m*=0.5).[26] Therefore, near half filling, $\tilde{f}$ may have the **Q**-displaced dispersion, too. A clue to discriminate these two candidates is the strongly electron-hole asymmetric amplitude of $\mathrm{Im}\Sigma^{\mathrm{nor}}$ (Fig.4). According to Eq.5, this asymmetry indicates that $\varepsilon_f$ is positive at $\mathbf{k}_{\mathrm{AN}}$. Since *t'* lowers $\varepsilon_c(\mathbf{k}_{\mathrm{AN}})$, $\varepsilon_f > 0$ is incompatible with the antiferromagnetic counterpart, while the composite fermion was shown to satisfy $\varepsilon_f > 0$ [25]. In this respect, $\tilde{f}$ is a stronger candidate. The composite fermions may form strongly-bound pairs because of their excitonic and dipole nature. Although they are incoherent Cooper pairs, they enhance the Cooper pairing of the quasiparticles[25,27]. An intriguing future issue is whether the hidden fermions under the dopant random potential form a large local gap as suggested experimentally[28] and enhance the superconducting gap by the proximity effect. Other issues also remain open; e.g., the relation of the present mechanism to proposed quantum criticalities and robustness of the present mechanism in multi-band systems.

A new picture for the high-$T_c$ superconducting mechanism emerges: The hybridization of quasiparticles with another fermionic excitation at the same time causes the pseudogap and boosts $T_c$. Though the pole cancellation of $\Sigma^{\mathrm{ano}}$ and $\Sigma^{\mathrm{nor}}$ in $G^{-1}$ makes the direct experimental detection of the hidden fermions difficult in the single-particle spectra, experiments for two-particle response functions (for example, optics[29,30]) could be re-interpreted in the light of the hidden-fermion mechanism.

**Methods**
**Numerical simulation on the microscopic model:** We study the 2D Hubbard model,

$$H_{\mathrm{Hubbard}} = -t \sum_{\langle ij \rangle \sigma} c_{i\sigma}^\dagger c_{j\sigma} - t' \sum_{\langle\langle ij \rangle\rangle \sigma} c_{i\sigma}^\dagger c_{j\sigma} - \mu \sum_{i\sigma} n_{i\sigma} + U \sum_i n_{i\uparrow} n_{i\downarrow}, \qquad (S1)$$

where $c_{i\sigma}(c_{i\sigma}^\dagger)$ annihilates (creates) an electron of spin σ at site *i* on a square lattice, and $n_{i\sigma} = c_{i\sigma}^\dagger c_{i\sigma}$. *t*(*t'*) denotes the (next-)nearest-neighbor transfer integral, *U* the onsite

Coulomb repulsion, and µ the chemical potential. Here the bare dispersion is given by

$\varepsilon(\mathbf{k}) = -2t(\cos k_x + \cos k_y) - 4t' \cos k_x \cos k_y - \mu.$  (S2)

We take $t=1$ as the unit of energy and adopt $t'=-0.2$ and $U=8$, which are reasonable for hole-doped cuprates.

We solve the model (S1) within the CDMFT[8], which gives the single-particle Green's function $\hat{\mathbf{G}}$, consisting of the normal part $G$ and the anomalous part $F$ in the Nambu representation for the superconducting state as

$$\hat{\mathbf{G}}(\mathbf{k},\omega) \equiv \begin{pmatrix} G(\mathbf{k},\omega) & F(\mathbf{k},\omega) \\ F(\mathbf{k},\omega) & -G(\mathbf{k},-\omega)^* \end{pmatrix} = \begin{pmatrix} \omega - \varepsilon(\mathbf{k}) - \Sigma^{nor}(\mathbf{k},\omega) & -\Sigma^{ano}(\mathbf{k},\omega) \\ -\Sigma^{ano}(\mathbf{k},\omega) & \omega + \varepsilon(\mathbf{k}) + \Sigma^{nor}(\mathbf{k},-\omega)^* \end{pmatrix}^{-1}.$$

Then the matrix inversion gives the normal Green's function

$G(\mathbf{k},\omega) = [\omega - \varepsilon(\mathbf{k}) - \Sigma^{nor}(\mathbf{k},\omega) - W(\mathbf{k},\omega)]^{-1}$

with

$W(\mathbf{k},\omega) = \Sigma^{ano}(\mathbf{k},\omega)^2 / [\omega + \varepsilon(\mathbf{k}) + \Sigma^{nor}(\mathbf{k},-\omega)^*],$

as Eqs.1 and 2. Note that in the limit of $T \to 0$, if a pole exist in $\Sigma^{ano}$ at $\omega^{ano}$, $\Sigma^{nor}$ is required to have a pole at the same energy $\omega^{nor} = \omega^{ano}$, since $W$ in Eq. 2 would have a pole of the order 2 at $\omega^{ano}$, which breaks causality unless $\Sigma^{nor}$ has a pole of the order 1 at the same energy.

Here, $G$ can also be rewritten as

$$G(\mathbf{k},\omega) = \frac{z(\mathbf{k},\omega)}{\omega - \tilde{\varepsilon}(\mathbf{k},\omega) z(\mathbf{k},\omega) - \tilde{W}(\mathbf{k},\omega)}$$

with

$$\tilde{W}(\mathbf{k},\omega) = \frac{[\Sigma^{ano}(\mathbf{k},\omega) z(\mathbf{k},\omega)]^2}{\omega + \tilde{\varepsilon}(\mathbf{k},\omega) z(\mathbf{k},\omega)} \quad \text{and} \quad \tilde{\varepsilon}(\mathbf{k},\omega) = \varepsilon(\mathbf{k}) + (\Sigma^{nor}(\mathbf{k},\omega) + \Sigma^{nor}(\mathbf{k},-\omega)^*)/2$$

from which the physical meaning of $z(\mathbf{k},\omega) = 1/[1 - (\Sigma^{nor}(\mathbf{k},\omega) - \Sigma^{nor}(\mathbf{k},-\omega)^*)/2\omega]$ as the renormalization factor (or quasiparticle residue) is clear. Namely, deviation of $z(\mathbf{k},\omega)$ from 1 measures electron correlation effects in the normal state. Furthermore, it connects the gap function $\Delta(\mathbf{k},\omega)$ to $\Sigma^{ano}(\mathbf{k},\omega)$ by the relation $\Sigma^{ano}(\mathbf{k},\omega) = \Delta(\mathbf{k},\omega)/z(\mathbf{k},\omega)$ equivalently to Eq.3.

With CDMFT, we map the model (S1) onto an effective Anderson model consisting of a 2-by-2 square cluster and bath sites. The effective model is solved by means of the exact-diagonalization method extended to finite temperatures[23], where we employ eight bath sites. This method enabled us to study the precise real-frequency electronic structures in the superconducting and the pseudogap phases on equal footing. Although the short-range spatial correlations within the 2-by-2 cluster are fully taken into account, the longer-range correlations are not incorporated in the calculation. This will be the reason why the calculated gap amplitude and $T_c$ look larger than those observed experimentally in cuprates. However, this will not modify the essence of the present article, as it has been shown that the 2-by-2 cluster DMFT provides results which are qualitatively consistent with various experiments and with larger-cluster calculations in both normal and superconducting states[7,9-20,22-24]. The spectra are obtained by replacing ω with ω+iη, where we employ η=0.05(1+ω$^2$) for |ω|<3 and η=0.5 for |ω|>3. The peak positions and the relative amplitudes are insensitive to the choice of η.

**Solution of the two-component fermion model:** The model, Eq.4, can be solved exactly and the normal (anomalous) part of Green's function, $G_c$ ($F_c$), for the $c$ electron is given by

$$G_c(\omega) = \frac{\omega^3 + \varepsilon_c \omega^2 - (\varepsilon_f^2 + D_f^2 + V^2)\omega + \varepsilon_f V^2 - \varepsilon_c(\varepsilon_f^2 + D_f^2)}{\det(\omega I - H_{\text{TCF}})},$$

$$F_c(\omega) = F_c^\dagger(\omega) = \frac{D_c(\omega^2 - \varepsilon_f^2 - D_f^2) - D_f V^2}{\det(\omega I - H_{\text{TCF}})}, \quad (S5)$$

$$\det(\omega I - H_{\text{TCF}}) = \omega^4 - (\varepsilon_c^2 + D_c^2 + \varepsilon_f^2 + D_f^2 + 2V^2)\omega^2 + V^4 + 2(-\varepsilon_c \varepsilon_f + D_c D_f)V^2 + (\varepsilon_c^2 + D_c^2)(\varepsilon_f^2 + D_f^2),$$

where we have omitted the **k** dependence for brevity. The normal and anomalous self-energies for the electron are defined by

$$\begin{pmatrix} \Sigma_c^{\text{nor}}(\omega) & \Sigma_c^{\text{ano}}(\omega) \\ \Sigma_c^{\text{ano}}(\omega) & -\Sigma_c^{\text{nor}}(-\omega)^* \end{pmatrix} = \begin{pmatrix} \omega - \varepsilon_c & 0 \\ 0 & \omega + \varepsilon_c \end{pmatrix} - \begin{pmatrix} G_c(\omega) & F_c(\omega) \\ F_c(\omega) & -G_c(-\omega)^* \end{pmatrix}^{-1}$$

$$= \begin{pmatrix} \omega - \varepsilon_c & 0 \\ 0 & \omega + \varepsilon_c \end{pmatrix} - \frac{1}{G_c(\omega)G_c(-\omega)^* + F_c(\omega)^2}\begin{pmatrix} G_c(-\omega)^* & F_c(\omega) \\ F_c(\omega) & -G_c(\omega) \end{pmatrix}.$$

(S6)

Substituting Eq.S5 into the above equation, we obtain Eq.5 in the main text. Here it is worth noting that Eq.S6 allows one to obtain $\Sigma^{\text{ano}}$ through the information on Green's functions $G$ and $F$, which would be measurable in experiments such as angle-resolved photoemission spectroscopy and optics.

Acknowledgment

S.S thanks A. Liebsch for fruitful discussions in making the program code for the numerical simulation, and R. Akashi for useful comments. This work was supported by Grant-in-Aid for Scientific Research (Grant No. 26800179, SS; No.22104010 and No. 22340090, MI), the Computational Materials Science Initiative (CMSI), HPCI Strategic Programs for Innovative Research (SPIRE), and RIKEN Advanced Institute for Computational Science (AICS) (Grant No. hp120043, hp120283, and hp130007), from MEXT, Japan.

# Supplementary Information

**Contribution to the superconducting gap:** From the Kramers-Kroenig relation, the gap function $\Delta$ satisfies

$$\mathrm{Re}\Delta(\mathbf{k}, \omega = 0) = \frac{2}{\pi} \int_0^\infty \frac{\mathrm{Im}\Delta(\mathbf{k},\omega')}{\omega'} d\omega' \qquad (S3)$$

In order to quantify the contribution to the gap $\mathrm{Re}\Delta(\mathbf{k},\omega=0)$ from $\mathrm{Im}\Delta$ in the each energy range $[0,\Omega]$, we define a function,

$$I(\mathbf{k},\Omega) \equiv \frac{2}{\pi \mathrm{Re}\Delta(\mathbf{k},\omega=0)} \int_0^\Omega \frac{\mathrm{Im}\Delta(\mathbf{k},\omega')}{\omega'} d\omega', \qquad (S4)$$

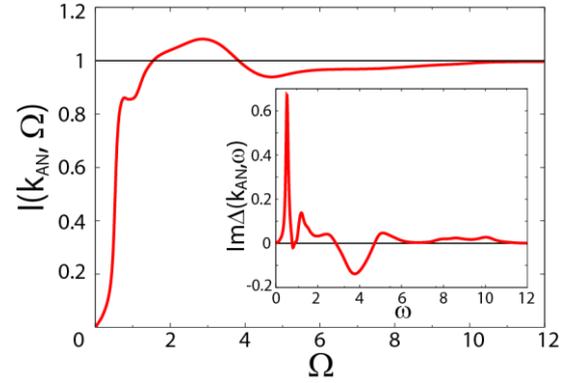

**Figure S1 | Integral weight contributing to the superconducting gap.** $I$ function (Eq.S4) at $\mathbf{k}=\mathbf{k}_{AN}$, calculated with the CDMFT for $x$=0.05 and $T$=0.01. Inset shows $\mathrm{Im}\Delta$ in the relevant energy range.

in a way similar to that used for $\Sigma^{ano}$ in Refs.10 and 12. We numerically evaluate this function for $\Delta$ calculated with the CDMFT (Fig.1c). The result shows that $I(\mathbf{k},\Omega)$ quickly arises around $\Omega$=0.5 indicating more than 80% of the weight comes from the pole of $\Delta$ at $\omega$=0.5, suggesting that the high-$T_c$ superconductivity is indeed caused by the low-energy peak in $\mathrm{Im}\Delta$.

**Further agreements between the CDMFT result and the solution of model Eq.4**

**[1] Shift of pole in $\Delta$ from $\Sigma^{ano}$**

The peak position of $\mathrm{Im}\Delta$ (at $\omega$~0.5 in Fig.1c) shifts from that of $\mathrm{Im}\Sigma^{ano}$ (at $\omega$~0.4 in Fig.1b), which is different from the standard ME theory, showing a novel role of the ω−dependent $\Sigma^{nor}$. A cancellation of poles similar to that between $\Sigma^{nor}$ and $W$ occurs in Eq.3, which results in a shift of the pole structure in $\Delta$ from that in $\Sigma^{ano}$: The poles of $\Delta$ are determined by zeros of ω-($\Sigma^{nor}$(k,ω) - $\Sigma^{nor}$(k,-ω)*)/2, which are located at a higher energy (|ω|~0.5) than the poles of $\Sigma^{ano}$(|ω|~0.4). In the two-component fermion model, on the other hand, substitution of Eq.5 into Eq.3 gives

$$\Delta(\mathbf{k},\omega) = D_c(\mathbf{k}) - V^2(D_f(\mathbf{k}) - D_c(\mathbf{k}))/(\omega^2 - \varepsilon_f(\mathbf{k})^2 - D_f(\mathbf{k})^2 - V^2), \qquad (S7)$$

which shows the shift of the poles from $\omega = \pm\sqrt{\varepsilon_f^2 + D_f^2}$ for $\Sigma^{ano}$ to $\omega = \pm\sqrt{\varepsilon_f^2 + D_f^2 + V^2}$ for $\Delta$, in accord with the CDMFT results in Figs.1b and c.

**[2] Zeros of $G$**

We also see a good correspondence between the CDMFT results for the Hubbard model and the two-component fermion model Eq.4 for the three zeros of $G$ (poles of $\Sigma^{nor}+W$) at low energy (|ω|<0.6). One originates from the particle-particle channel that generates the superconducting gap. This is at ω~0.15 in Fig.3a and corresponds to a zero of $G_c$ for the model Eq.4, whose energy is obtained by solving the zero of the numerator of $G_c$ in Eq.S5 (given by one of the three roots of the cubic equation), and reduces to $-\varepsilon_c$ in the limit $V$=0. This is distinguished from the other two peaks at ω~ ± 0.5 in Fig.3a, corresponding to the other two zeros of $G_c$ in Eq.S5. The latter zeros approach $\omega = \pm\sqrt{\varepsilon_f(\mathbf{k})^2 + D_f(\mathbf{k})^2}$ for $V \to 0$, indicating the origin from the particle-hole channel generating the pseudogap.

**Fitting with the two-component fermion model in Fig. 4:**

In the two-component fermion model (Eq.4), $\Sigma_c$ is generated by Green's functions of $f$ for $V$=0, given by $G_{f,0}^{nor}=(\omega+\varepsilon_f)/(\omega^2-\varepsilon_f^2-D_f^2)$ and $G_{f,0}^{ano}=D_f/(\omega^2-\varepsilon_f^2-D_f^2)$, through the mutual hybridization $V$, i.e., $\Sigma_c^{nor}=V^2 G_{f,0}^{nor}$ and $\Sigma_c^{ano}=D_c-V^2 G_{f,0}^{ano}$. Hence, the poles of $\Sigma_c$ and $G_{f,0}$ are the same[28]. This helps to clarify the character of the $f$ fermion. Namely, by calculating the **k**-dependent self-energy with the CDMFT, we can deduce the bare dispersion of $f$. The **k**-dependent quantities of CDMFT in Figs.4a-d have been calculated through an interpolation scheme based on Green's function[21], which allows us to obtain the **k**-dependent positions of the peaks in Im$\Sigma^{ano}$.

In fitting the CDMFT results with Eq.5, we have employed $\varepsilon_f(\mathbf{k}) = \varepsilon_c(\mathbf{k}+\mathbf{Q}) - \mu_f$ with $\varepsilon_c(\mathbf{k}) = z(\mathbf{k})(\varepsilon(\mathbf{k}) + \mu)$, where the renormalization factor $z(\mathbf{k})$ is obtained with the CDMFT. In a typical case at $T$=0.01 and $x$=0.04, the CDMFT result shows $z(\mathbf{k})$ is 0.114 at $\mathbf{k}=(\pi,0)$ and 0.105 at $\mathbf{k}=(\pi/2, \pi/2)$. We take the *d*-wave form for the order parameters; $D_{c,f}(\mathbf{k}) = \frac{D_{c,f}^0}{2}(\cos k_y - \cos k_x)$.

Exploiting the fact that both $\Sigma^{\text{nor}}$ and $\Sigma^{\text{ano}}$ have the poles at $\omega = \pm\sqrt{\varepsilon_f(\mathbf{k})^2 + D_f(\mathbf{k})^2}$ (Eq.5), we determine the parameter values used in Fig.4e-h: $\mu_f$ is determined by referring the peak position of the CDMFT result for $\text{Im}\Sigma^{\text{nor}}$ (Fig.4a) at $\mathbf{k}=(\pi/2,\pi/2)$, where the anomalous part vanishes. Then $D_f^0$ is determined by the peak position at $(\pi,0)$. $V$ is determined by fitting the peak amplitude of $\text{Im}\Sigma^{\text{ano}}$ at $\mathbf{k}=(\pi,0)$. Notice that the values of $\varepsilon_c$ and $D_c^0$ are irrelevant to $\text{Im}\Sigma_c$ as far as we take $D_c^0$ to be real. We employ the same broadening factor $\eta$ as that used in the CDMFT.

**Construction of the orthogonal excitation $\tilde{f}$:** Here we construct a local operator $\tilde{f}_{i\sigma}$ which is approximately orthogonal to the electron operator $c_{i\sigma}$. Taking account of the different local states in the single-orbital Hubbard model, we consider a linear combination,

$\tilde{f}_{i\sigma} \equiv A c_{i\sigma} + B c_{i\sigma} n_{i\bar{\sigma}}$ (*A* and *B* are the constants determined below),

of the two possible local operators, $c_{i\sigma}$ and $c_{i\sigma} n_{i\bar{\sigma}}$. The anticommutation relations of $\tilde{f}$ with itself and with *c* are given by

$$\begin{aligned}
\{c_\sigma, \tilde{f}_\sigma\} &= \{\tilde{f}_\sigma, \tilde{f}_\sigma\} = \{\tilde{f}_\sigma, \tilde{f}_{\bar{\sigma}}\} = 0, \\
\{c_\sigma, \tilde{f}_{\bar{\sigma}}\} &= B c_\sigma c_{\bar{\sigma}}, \\
\{c_\sigma^\dagger, \tilde{f}_\sigma\} &= A + B n_{\bar{\sigma}}, \\
\{c_\sigma^\dagger, \tilde{f}_{\bar{\sigma}}\} &= B c_{\bar{\sigma}} c_\sigma^\dagger, \\
\{\tilde{f}_\sigma^\dagger, \tilde{f}_\sigma\} &= |A|^2 + \left(A\bar{B} + \bar{A}B + |B|^2\right) n_{\bar{\sigma}}, \\
\{\tilde{f}_\sigma^\dagger, \tilde{f}_{\bar{\sigma}}\} &= \left(|A+B|^2 - |A|^2\right) c_{\bar{\sigma}} c_\sigma^\dagger,
\end{aligned} \quad \text{(S8)}$$

and the Hermitian conjugates of these (here we omitted the site index for brevity). We would like to determine *A* and *B* to make $\tilde{f}$ orthonormal, i.e., the right hand side of Eq.S8 to be zero, except for the fifth line to be 1. Although it is impossible to satisfy all of them rigorously, it is possible to determine *A* and *B* to satisfy such anticommutation relations in an approximate way, i.e., at a level of the thermal average $\langle ... \rangle \equiv \frac{\text{Tr}[e^{-\beta H}...]}{\text{Tr}\,e^{-\beta H}}$, where $\beta$ denotes the inverse temperature. To see this, we first take the thermal average of the second line in Eq.S8; it becomes

$$\langle\{c_\sigma,\tilde{f}_{\bar\sigma}\}\rangle = B\langle c_\sigma c_{\bar\sigma}\rangle = 0$$

because we are considering the *d*-wave superconductivity. Similarly, the fourth and the last lines in Eq.S8 become

$$\langle\{c_\sigma^\dagger,\tilde{f}_{\bar\sigma}\}\rangle = B\langle c_{\bar\sigma} c_\sigma^\dagger\rangle = 0,$$

$$\langle\{\tilde{f}_\sigma^\dagger,\tilde{f}_{\bar\sigma}\}\rangle = \left(|A+B|^2 - |A|^2\right)\langle c_{\bar\sigma} c_\sigma^\dagger\rangle = 0$$

because of the spin conservation. Then we require

$$\langle\{c_\sigma^\dagger,\tilde{f}_\sigma\}\rangle = A + B\langle n_{\bar\sigma}\rangle = 0$$

and

$$\langle\{\tilde{f}_\sigma^\dagger,\tilde{f}_\sigma\}\rangle = |A|^2 + \left(A\bar{B} + \bar{A}B + |B|^2\right)\langle n_{\bar\sigma}\rangle = 1.$$

These two equations are satisfied for

$$A = \sqrt{\frac{m}{1-m}}, \quad B = -\sqrt{\frac{1}{m(1-m)}}$$

with $m \equiv \langle n_{\bar\sigma}\rangle$.

With thus-determined $\tilde{f}_{i\sigma} \equiv \sqrt{\frac{m}{1-m}}c_{i\sigma} - \sqrt{\frac{1}{m(1-m)}}c_{i\sigma}n_{i\bar\sigma}$, the Hubbard interaction term can be recast into

$$Un_\uparrow n_\downarrow = -U(1-\gamma)\frac{\sqrt{m(1-m)}}{4}\sum_\sigma\left(c_\sigma^\dagger \tilde{f}_\sigma + \tilde{f}_\sigma^\dagger c_\sigma\right) + U\gamma\frac{m(1-m)}{2(1-2m)}\sum_\sigma \tilde{f}_\sigma^\dagger \tilde{f}_\sigma + U\frac{m}{2}\left(1-\gamma\frac{1-m}{1-2m}\right)\sum_\sigma n_\sigma,$$

where γ is a number which controls the shift of the chemical potential of *c* through the third term on the right-hand side. The first (second) term will give $V$ ($\mu_f$) term in the model, Eq.4.

15